\documentclass[prl,showpacs,twocolumn,aps]{revtex4}

\usepackage{amsmath}
\usepackage{amssymb}
\usepackage{latexsym}
\usepackage{amsfonts}
\usepackage{epsfig}
\usepackage{psfrag}
\usepackage{graphicx}

\usepackage{color}

\newcommand{\ket}[1]{\left|#1\right>}
\newcommand{\bra}[1]{\left<#1\right|}   

\newcommand{\nn}{\nonumber\\}

\newcommand{\f}[1]{\mbox{\boldmath$#1$}}
\newcommand{\fk}[1]{\mbox{\boldmath$\scriptstyle#1$}}

\newcommand{\na}{\mbox{\boldmath$\nabla$}}
\newcommand{\bea}{\begin{eqnarray}}
\newcommand{\ea}{\end{eqnarray}}
\newcommand{\eea}{\end{eqnarray}}
\newcommand{\ord}{\,{\cal O}}
\newcommand{\li}{\,\widehat{\cal L}}
\newcommand{\tr}{\,{\rm Tr}}

\begin{document}

\title{Sauter-Schwinger like tunneling in tilted Bose-Hubbard lattices 
in the Mott phase} 

\author{Friedemann Queisser$^1$, Patrick Navez$^2$, and 
Ralf Sch\"utzhold$^{1,*}$}

\affiliation{
$^1$Fakult\"at f\"ur Physik, Universit\"at Duisburg-Essen, 
Lotharstrasse 1, 47057 Duisburg, Germany
\\
$^2$Institut f\"ur Theoretische Physik, 
Technische Universit\"at Dresden, 01062 Dresden, Germany}
 
\date{\today}

\begin{abstract}
We study the Mott phase of the Bose-Hubbard model on a tilted lattice. 
On the (Gutzwiller) mean-field level, the tilt has no effect -- but 
quantum fluctuations entail particle-hole pair creation via tunneling. 
For small potential gradients (long-wavelength limit), we derive a 
quantitative analogy to the Sauter-Schwinger effect, i.e., electron-positron 
pair creation out of the vacuum by an electric field.
For large tilts, we obtain resonant tunneling related to Bloch oscillations.
\end{abstract}


\maketitle

\paragraph{Introduction.}

\begin{figure}[h]
\includegraphics[width=\columnwidth]{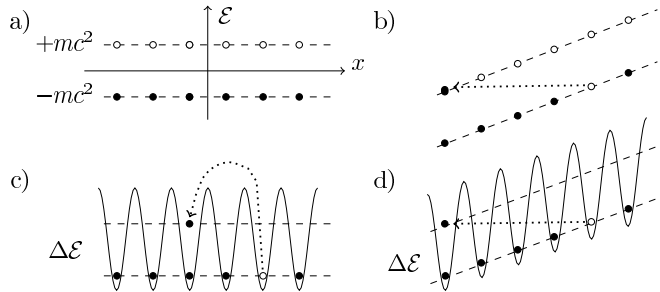}
\psfrag{test}{$\Delta\cal E$}
\caption{Sketch of the analogy: 
a) Dirac sea for $E=0$, 
b) Sauter-Schwinger tunneling for $E\neq0$, 
c) Mott state with energy gap $\Delta\cal E$, 
d) tunneling in tilted lattice.}\label{analog}
\label{figure}
\end{figure}

There are many striking analogies between apparently very different areas 
in physics, which help us to unify our understanding of nature -- 
as R.~Feynman said: {\em The same equations have the same solutions.}
In this Letter, we establish such a quantitative analogy between 
high-field science and ultra-cold atoms in optical lattices. 
The Sauter-Schwinger effect \cite{Schwinger1,Schwinger2,Schwinger3} %\cite{S31,S51,D09,BI70,SGD08,K65} 
predicts that an extremely strong electric field may create 
electron-positron pairs out of the QED vacuum.
As an intuitive picture, one can envisage this effect as tunneling of an 
electron from the Dirac sea into the positive continuum, leaving behind 
a hole (i.e., positron), see Fig. \ref{analog}.
Unfortunately, this prediction has not been experimentally verified yet 
since it is hard to generate sufficiently strong electric fields in large 
enough space-time regions.
As we shall demonstrate below, the same equations govern particle-hole 
pair creation in a Mott insulator via tunneling due to a small potential 
gradient. 
The Mott insulator state we are considering can be generated by bosonic 
atoms in an optical lattice \cite{BoseHubbard,Bloch} for example, which repel each other and 
thereby become effectively pinned to the lattice sites (zero mobility).
The benefits of this quantitative analogy are two-fold.
On the one hand, we can apply our knowledge of the Sauter-Schwinger effect 
\cite{Schwinger1,Schwinger2} to atoms in optical lattices and understand them better in this way.
On the other hand, atoms in optical lattices may provide (via this analogy) 
an experimental approach to the so far unobserved Sauter-Schwinger effect.
%\newpage 
\paragraph{Sauter-Schwinger effect.}
Let us start with the Dirac equation describing relativistic electrons 
($\hbar=c=1$)
\begin{equation}
\label{Dirac}
\gamma^\mu(i\partial_\mu-qA_\mu)\Psi-m\Psi=0
\,,
\end{equation}
with charge $q$ and mass $m$, propagating in an electromagnetic field $A_\mu$.
An electric field $\f{E}$ can be encoded in the scalar potential 
$\f{E}=\na\Phi$ or be generated by the vector potential 
$\f{E}=\partial_t\f{A}$.
Using the standard representation of the Dirac matrices $\gamma^\mu$ 
via the Pauli matrices $\f{\sigma}$ and splitting the bi-spinor $\Psi$ 
into upper and lower components $\Psi=(\psi_+,\psi_-)$, 
the ``square'' of Eq.~(\ref{Dirac}) yields 
\begin{equation}
\label{Klein-Gordon}
\left[
\left(i\partial_t-q\Phi\right)^2
-\left(i\na+q\f{A}\right)^2
-m^2
\right]
\psi_\pm
=
iq\f{E}\cdot\f{\sigma}\,\psi_\mp
\,.
\end{equation}
For sub-critical electric fields $m^2\gg qE$, we may neglect the 
spin-dependent term on r.h.s.\ and obtain the Klein-Fock-Gordon 
equation for both spinors $\psi_\pm$ separately.
If the electric field is purely time-dependent $\f{E}=E(t)\f{e}_x$, 
the temporal gauge $\Phi=0$ and $\f{E}=\partial_t\f{A}$ is most 
convenient: 
After a spatial Fourier transform, we find that each mode $\f{k}$ 
corresponds to a harmonic oscillator 
\bea
\label{scattering}
\left[
\frac{d^2}{dt^2}
+[k_x+qA(t)]^2
+\f{k}_\perp^2
+m^2
\right]
\psi_{\fk{k}}
=0
\,,
\ea
with a time-dependent potential.
Replacing $t$ by $x$, the above equation is equivalent to a one-dimensional 
Schr\"odinger scattering problem.  
A solution which initially behaves as 
$\psi_{\fk{k}}^{\rm in}(t)\propto\exp\{-i\omega_{\fk{k}}t\}$
will finally evolve into a mixture of positive and negative frequencies 
$\psi_{\fk{k}}^{\rm out}(t)\propto
\alpha_{\fk{k}}\exp\{-i\omega_{\fk{k}}t\}+
\beta_{\fk{k}}\exp\{+i\omega_{\fk{k}}t\}$.
The Bogoliubov coefficient $\beta_{\fk{k}}$ is related to the reflection 
amplitude in one-dimensional scattering theory and yields the probability 
for electron-positron pair creation. 
For slowly varying fields $E(t)$ with $m^2\gg qE$, 
it can be derived via the WKB approximation.
Since Eq.~(\ref{scattering}) corresponds to scattering above the barrier,
the turning points are not real but complex. 
Hence the reflection (i.e., pair creation) probability is exponentially 
suppressed \cite{Schwinger1} demonstrating that it is a non-perturbative effect 
\bea
\label{probability}
P^{e^+e^-}_{\fk{k},-\fk{k}}
=
|\beta^2_{\fk{k}}|
\propto
\exp\left\{-\pi\,\frac{m^2+\f{k}_\perp^2}{qE}\right\}
\,,
\ea
where $E$ denotes the maximum electric field.

\paragraph{Bose-Hubbard model.}

We consider the Hamiltonian
\bea
\label{Hamiltonian}
\hat H
=
-\frac{J}{Z}\sum_{\mu,\nu} T_{\mu\nu} \hat a^\dagger_\mu \hat a_\nu
+\frac{U}{2}\sum_{\mu} (\hat a^\dagger_\mu)^2\hat a_\mu^2
%\hat n_\mu(\hat n_\mu-1)
+\sum_{\mu} V_\mu \hat n_\mu
\,,
%\nn
\ea
with the creation and annihilation operators $\hat a^\dagger_\mu$ and 
$\hat a_\nu$ at the lattice sites $\mu$ and $\nu$, respectively. 
The tunneling matrix $T_{\mu\nu}\in\{0,1\}$ encodes the lattice structure
and $J$ denotes the hopping rate. 
The number of tunneling neighbors at any given site $\mu$ yields 
the coordination number $Z=\sum_{\nu} T_{\mu\nu}\in\mathbb N$.
Finally, $U$ is the on-site interaction and $V_\mu$ denotes the 
on-site potential with $\hat n_\mu=\hat a^\dagger_\mu \hat a_\mu$. 
This potential $V_\mu$ at lattice position $\f{r}_\mu$
will be analogous to the electric field $\f{E}$ via 
$V_\mu(t)\leftrightarrow q\f{r}_\mu\cdot\f{E}(t)$. 

We assume an average filling of one boson per site 
$\langle\hat n_\mu\rangle=1$.
In order to have well-defined initial and final states, we envisage the 
following sequence:
Initially, we have $J=V_\mu=0$ where the ground state simply reads 
$\ket{\Psi_{\rm in}}=\prod_\mu\ket{1}_\mu$ (Mott insulator).
Then we switch on the hopping rate $J$ adiabatically 
(i.e., slow compared to the energy gap of the Mott insulator) 
such that we stay in the ground state. 
Next we introduce the lattice potential $V_\mu(t)$ which will be analogous 
to the electric field and enables particle-hole creation via tunneling. 
Again, we should do this slowly in order to avoid dynamical excitations
\cite{Schwinger2}.
Finally, we reverse this process and slowly switch off first $V_\mu(t)$ 
and then $J$. 
Thus the final ground state again reads $\prod_\mu\ket{1}_\mu$.
If the final quantum state does not have exactly one particle per site
$\ket{\Psi_{\rm out}}\neq\prod_\mu\ket{1}_\mu$
we have a signature of particle-hole pair creation. 

In order to calculate this effect, we proceed along the lines of \cite{NavSch}
%\cite{NS10} 
and consider the time evolution of the density operator $\hat\rho$ 
of the whole lattice and introduce Liouville 
super-operators $\li_\mu$ and $\li_{\mu \nu}$ via 
\bea
\label{Liouville}
i \partial_t \hat\rho = \left[\hat H,\hat\rho\right] 
= 
\frac{1}{Z}\sum_{\mu,\nu} \li_{\mu \nu}\hat\rho
+
\sum_\mu\li_\mu\hat\rho
\,. 
\ea
As the next step, we derive the reduced density matrices for one 
lattice site $\hat\rho_\mu=\tr_{\not\mu}\{\hat\rho\}$ via averaging 
(tracing) over all other sites $\nu\neq\mu$ and similarly for 
two and more sites 
$\hat\rho_{\mu\nu}=\tr_{\not\mu\not\nu}\{\hat\rho\}$ etc.
Then we separate the correlated parts via  
$\hat\rho_{\mu\nu}=\hat\rho_{\mu\nu}^{\rm corr}+\hat\rho_{\mu}\hat\rho_{\nu}$
for two sites, as well as 
$\hat\rho_{\mu\nu\lambda}=\hat\rho_{\mu\nu\lambda}^{\rm corr}+
\hat\rho_{\mu\nu}^{\rm corr}\hat\rho_{\lambda}+
\hat\rho_{\mu\lambda}^{\rm corr}\hat\rho_{\nu}+
\hat\rho_{\nu\lambda}^{\rm corr}\hat\rho_{\mu}+
\hat\rho_{\mu}\hat\rho_{\nu}\hat\rho_{\lambda}$
etc.
From Eq.~(\ref{Liouville}), we obtain the effective evolution equation 
for one lattice site $\mu$ via tracing over the rest $\nu\neq\mu$
\bea
\label{one-site}
i\partial_t\hat\rho_{\mu}
=
\frac{1}{Z}
\sum_{\kappa\neq\mu}\tr_{\kappa}\left\{
\li^S_{\mu \kappa}
(\hat\rho^{\rm corr}_{\mu \kappa}+\hat\rho_\mu \hat\rho_\kappa)\right\}
+
\li_\mu  \hat\rho_{\mu}
\,,
\eea
where $\li_{\mu \nu}^S=\li_{\mu \nu}+\li_{\nu \mu}$.
In order to employ a controlled analytic approach, we consider the limit
of large coordination numbers $Z\gg1$.
In this limit, the correlations obey the following hierarchy \cite{hierarchy,NavSch}.
While the one-site density matrix $\hat\rho_\mu=\ord(Z^0)$ has entries of
order one $\tr_\mu\{\hat\rho_\mu\}=1$, the two-site correlations are 
suppressed via $\hat\rho_{\mu\nu}^{\rm corr}=\ord(1/Z)$, the three-site 
correlations via $\hat\rho_{\mu\nu\lambda}^{\rm corr}=\ord(1/Z^{2})$
and so on. 
Thus, Eq.~(\ref{one-site}) reproduces the Gutzwiller ansatz to lowest order 
in $1/Z$, which reads 
$\hat\rho_{\mu}=\ket{1}_\mu\!\bra{1}+\ord(1/Z)=\hat\rho_{\mu}^0+\ord(1/Z)$
in the Mott state. 
Inserting this result and neglecting terms of order $\ord(1/Z^2)$, 
we get for two sites \cite{NavSch}
\bea
\label{two-sites}
i \partial_t \hat\rho^{\rm corr}_{\mu \nu}
&=&
\frac1Z
\sum_{\kappa\not=\mu,\nu} 
\tr_{\kappa}
\left\{
\li^S_{\mu\kappa}\hat\rho^{\rm corr}_{\nu\kappa}\hat\rho_{\mu}^0
+\li^S_{\nu\kappa}\hat\rho^{\rm corr}_{\mu\kappa}\hat\rho_{\nu}^0
\right\}
+
\nn
&&+
\left(\li_\mu+\li_\nu\right)\hat\rho^{\rm corr}_{\mu\nu}
+
\frac1Z\li_{\mu\nu}^S\hat\rho_\mu^0\hat\rho_\nu^0
\,.
\eea
Formally, this is an evolution equation for an infinite dimensional matrix
$\hat\rho^{\rm corr}_{\mu \nu}$, but fortunately it is sufficient to 
consider four elements only. 
Introducing local particle and hole operators 
$\hat p_\mu=\ket{1}_\mu\!\bra{2}$ 
and 
$\hat h_\mu=\ket{0}_\mu\!\bra{1}$,  
we find that their correlation functions 
$f^{11}_{\mu\nu}=\langle\hat h_\mu^\dagger\hat h_\nu\rangle$,  
%
%$f^{11}_{\mu\nu}=\langle\hat h_\mu^\dagger\hat h_\nu\rangle
%=\tr\{\hat\rho\,\hat h_\mu^\dagger\hat h_\nu\}
%=\tr\{\hat\rho^{\rm corr}_{\mu\nu}\hat h_\mu^\dagger\hat h_\nu\}$,
%as well as,   
$f^{12}_{\mu\nu}=\langle\hat h_\mu^\dagger\hat p_\nu\rangle$, 
$f^{21}_{\mu\nu}=\langle\hat p_\mu^\dagger\hat h_\nu\rangle$,
and 
$f^{22}_{\mu\nu}=\langle\hat p_\mu^\dagger\hat p_\nu\rangle$, 
%
%
%\bea
%f^{11}_{\mu\nu}
%&=&
%\langle\hat h_\mu^\dagger\hat h_\nu\rangle
%=
%\tr\{\hat\rho\,\hat h_\mu^\dagger\hat h_\nu\}
%=
%\tr\{\hat\rho^{\rm corr}_{\mu\nu}\hat h_\mu^\dagger\hat h_\nu\}
%\,,
%\nn
%f^{12}_{\mu\nu}
%&=&
%\langle\hat h_\mu^\dagger\hat p_\nu\rangle
%\,,\;
%f^{21}_{\mu\nu}=\langle\hat p_\mu^\dagger\hat h_\nu\rangle
%\,,\;
%f^{22}_{\mu\nu}=\langle\hat p_\mu^\dagger\hat p_\nu\rangle
%\,,
%\ea
%
obey for $\nu\neq\mu$ a closed linear system of equations
\begin{eqnarray}
\left(i\partial_t+V_\mu-V_\nu-U\right)f^{12}_{\mu\nu}
=
-\frac{J\sqrt{2}}{Z}T_{\mu\nu}-
\nn
-
\frac{J}{Z}\sum_{\kappa\neq \mu,\nu}T_{\mu\kappa}
\left[3 f^{12}_{\kappa\nu}+\sqrt{2}f^{22}_{\kappa\nu}+
\sqrt{2}f^{11}_{\kappa\nu}\right]
\,,
\nn
\label{f11}
\left(i\partial_t+V_\mu-V_\nu\right) 
f^{11}_{\mu\nu}
=
\left(i\partial_t+V_\mu-V_\nu\right) 
f^{22}_{\mu\nu}
=
\nn
-\frac{\sqrt{2}J}{Z}\sum_{\kappa\neq\mu,\nu}T_{\mu\kappa}\left(
f_{\kappa\nu}^{21}-f_{\kappa\nu}^{12}\right)
\,,
\quad
\end{eqnarray}
together with the symmetry $f_{\mu\nu}^{12}=(f_{\nu\mu}^{21})^*$.
Apart from this trivial relation, we find an effective particle hole
symmetry $f^{11}_{\mu\nu}=f^{22}_{\mu\nu}$
(to first order in $1/Z$).

\paragraph{The Analogy.}

Eqs.~(\ref{f11}) provide a complete set of equations whose solution yields 
the number of particle-hole pairs created by the lattice tilt.
However, instead of solving them directly, we make the following {\em trick}. 
It turns out that we may recover (i.e., factorize) Eqs.~(\ref{f11}) 
to first order in $1/Z$ if we assume the following {\em effective} 
linear equations of the operators $\hat{h}_\mu$ and $\hat{p}_\mu$ 
\begin{eqnarray}
\label{effective}
\left[i\partial_t-V_\mu-\frac{U}{2}\right]\hat{p}_\mu
&=&
-\frac{J}{Z}\sum_\nu T_{\mu\nu}
\left[\frac32\,\hat{p}_\nu+\sqrt{2}\,\hat{h}_\nu\right]
\,,
\nn
\left[i\partial_t-V_\mu+\frac{U}{2}\right]\hat{h}_\mu
&=&
\frac{J}{Z}\sum_\nu T_{\mu\nu}
\left[\frac32\,\hat{h}_\nu+\sqrt{2}\,\hat{p}_\nu\right]
\,,
\end{eqnarray}
and exploit the initial conditions 
$\langle\hat h_\mu^\dagger\hat h_\nu\rangle_0=\delta_{\mu\nu}$
and 
$\langle\hat h_\mu^\dagger\hat p_\nu\rangle_0=
\langle\hat p_\mu^\dagger\hat h_\nu\rangle_0=
\langle\hat p_\mu^\dagger\hat p_\nu\rangle_0=0$
in the Mott state. 
Note that these are effective equations of motion -- 
they are {\em not} obtained by 
$i\partial_t\hat{h}_\mu=[\hat{h}_\mu,\hat H]$ etc.
Nevertheless, the two-point functions in Eqs.~(\ref{f11}) behave 
{\em as if} the operators $\hat{h}_\mu$ and $\hat{p}_\mu$ evolve
according to (\ref{effective}).
In this sense, solving Eqs.~(\ref{effective}) yields the correct physics.  

As the next step, we consider the (continuum) limit of large length scales 
%much larger than the lattice spacing (continuum limit) 
and approximate the tunneling matrix 
\bea
\frac{1}{Z}\sum_\nu T_{\mu\nu} \hat{h}_\nu(t)
\approx
\left[1+\xi\na^2+\ord(\na^4)\right]\hat h(t,\f{r})
\ea
by a long-wavelength expansion, where $\xi=\ord(1/Z)$ is the stiffness. 
Introducing effective scalar fields 
$\hat\phi_\pm(t,\f{r})=\hat h(t,\f{r})\pm\hat p(t,\f{r})$, 
%
%\bea
%\label{phi}
%\hat\phi_\pm(t,\f{r})=\hat h(t,\f{r})\pm\hat p(t,\f{r})
%\ea
%
we thus find the set of equations 
\bea
\label{pm}
\left(i\partial_t-V\right)\hat\phi_\pm=
\frac12
\left[\left(3\mp2\sqrt2\right)
{\cal D}-U\right]
\hat\phi_\mp
\ea
where ${\cal D}=J[1+\xi\na^2+\ord(\na^4)]$.
As we shall see later, a very small tilt of the lattice will only induce a 
significant probability for pair creation if we are close to the critical 
point: $U=J(3+2\sqrt2)+\varepsilon$ with $0<\varepsilon\ll U$. 
In this case, the derivative part $\xi\na^2$ can be neglected in the term 
$(3-2\sqrt2){\cal D}$ and the coupled equations (\ref{pm}) simplify to 
\bea
\label{scalar-field}
\left(i\partial_t-V\right)^2\hat\phi_+
=
\left[m^2_{\rm eff}c^4_{\rm eff}
-c^2_{\rm eff}\na^2+\ord(\na^4)\right]\hat\phi_+
\,.
\ea
Up to short-distance corrections $\ord(\na^4)$, we obtain the analogue of the 
Klein-Fock-Gordon equation (\ref{Klein-Gordon}) 
with an effective light velocity $c^2_{\rm eff}=\xi(3JU-J^2)/2$ 
\footnote{The expression for $c_{\rm eff}$ in \cite{NavSch} contained a typo.}
and an effective mass $m_{\rm eff}$ which is related to the energy gap 
\bea
\label{gap}
\Delta{\cal E}=\sqrt{J^2-6JU+U^2}=2m_{\rm eff}c^2_{\rm eff}
\ea
of the Mott insulator (in analogy to the energy gap $2m_ec^2$ 
between the Dirac sea and the positive energy solutions). 
%
%
%Note that this is formally the same relation as for the energy gap $2m_ec^2$ 
%separating the Dirac sea from the positive energy solutions for electrons.  
%
Even if we are not close to the critical point, the above equation 
(\ref{scalar-field}) should remain a good approximation since the ratio 
$(3-2\sqrt{2})/(3+2\sqrt{2})\approx 0.03\ll1$
of the two numerical factors in (\ref{pm}) happens to be quite small.

\paragraph{Particle-hole pair creation.}

For potentials of the form 
%
%which correspond to spatially homogeneous electric fields 
%
$V_\mu(t)=q\f{r}_\mu\cdot\f{E}(t)$, 
%
%$\f{E}(t)$, 
%
%The effective Klein-Fock-Gordon equation (\ref{scalar-field}) is valid for
%arbitrary potentials $V_\mu(t)\approx V(t,\f{r})$ but restricted to small
%momenta (large distances).
%
%In the following, we focus on potentials $V_\mu(t)=q\f{r}_\mu\cdot\f{E}(t)$
%which correspond to a spatially homogeneous electric field $\f{E}(t)$.
%
%In this case, 
%
it is convenient to apply a Fourier transform after shifting 
the potential $V_\mu(t)$ into the phase 
\bea
\hat{h}_\mu(t)e^{i\int_0^t dt'\,V_\mu(t')}
%\hat{h}_\mu(t)\exp\left\{i\int\limits_0^t dt'\,V_\mu(t')\right\}
=
\sum\limits_{\fk{k}}
\hat h_{\fk{k}}(t)
e^{i\fk{k}\cdot\fk{r}_\mu}
%\exp\{i\f{k}\cdot\f{r}_\mu\}
\,,
\ea
and analogously for $\hat{p}_\mu(t)$. 
Then Eqs.~(\ref{effective}) become 
\begin{eqnarray}
\label{eff-k}
i\partial_t \hat h_{\fk{k}}
&=&
+\frac12\left[3JT_{\fk{k}}-U\right]\hat h_{\fk{k}}
+\sqrt{2}\,JT_{\fk{k}}\,\hat p_{\fk{k}}
\,,
\nn
i\partial_t \hat p_{\fk{k}}
&=&
-\frac12\left[3JT_{\fk{k}}-U\right]\hat p_{\fk{k}}
-\sqrt{2}\,JT_{\fk{k}}\,\hat h_{\fk{k}}
\,,
\end{eqnarray}
where the $T_{\fk{k}}$ are now time-dependent 
\bea
\label{T_k}
\frac{Z}{N}
\sum\limits_{\fk{k}}
T_{\fk{k}}
e^{i\fk{k}\cdot(\fk{r}_\mu-\fk{r}_\nu)}
=
T_{\mu\nu}
\,
e^{-i\int\limits_0^t dt'\,[V_\mu(t')-V_\nu(t')]}
\,.
\ea
Inserting 
$V_\mu(t)=q\f{r}_\mu\cdot\f{E}(t)=\partial_t[q\f{r}_\mu\cdot\f{A}(t)]$, 
we see that the time dependence of $T_{\fk{k}}$ can be understood as 
replacing $\f{k}$ by $\f{k}+q\f{A}(t)$.  
%where $\f{A}(t)$ is the vector potential $\f{E}(t)=\partial_t\f{A}(t)$.  
%
This is also known as Peierls substitution and is completely analogous 
to the gauge transformation in electrodynamics discussed between 
Eqs.~(\ref{Dirac}) and (\ref{Klein-Gordon}).
It  also underlies the well-known Bloch oscillations. 

The most general solution of (\ref{eff-k}) can be written as 
\bea
\hat h_{\fk{k}}(t)
&=&
f^+_{\fk{k}}(t)\hat A_{\fk{k}}+
f^-_{\fk{k}}(t)\hat B_{\fk{k}}
\,,
\nn
\hat p_{\fk{k}}(t)
&=&
g^+_{\fk{k}}(t)\hat A_{\fk{k}}+
g^-_{\fk{k}}(t)\hat B_{\fk{k}}
\,,
\ea
with suitable functions $f^\pm_{\fk{k}}(t)$ and $g^\pm_{\fk{k}}(t)$ and 
operators $\hat A_{\fk{k}}$ and $\hat B_{\fk{k}}$. 
In the initial stationary regime where $J=V_\mu=0$, we have  
$\hat h_{\fk{k}}(t)=\hat h_{\fk{k}}^{\rm in}\exp\{iUt/2\}$ and 
$\hat p_{\fk{k}}(t)=\hat p_{\fk{k}}^{\rm in}\exp\{-iUt/2\}$. 
Thus, w.l.o.g.\ we may identify 
$\hat h_{\fk{k}}^{\rm in}$ with $\hat A_{\fk{k}}$ and 
$\hat p_{\fk{k}}^{\rm in}$ with $\hat B_{\fk{k}}$.
This implies $f^+_{\fk{k}}(t)\propto\exp\{iUt/2\}$ and
$g^-_{\fk{k}}(t)\propto\exp\{-iUt/2\}$ initially with 
%while the other two vanish initially 
$g^+_{\rm in}=f^-_{\rm in}=0$.
During the time evolution according to (\ref{eff-k}), however, 
$f^\pm_{\fk{k}}(t)$ and $g^\pm_{\fk{k}}(t)$ will mix and 
positive and negative frequencies will not stay separated -- 
as discussed after Eq.~(\ref{scattering}). 
In the final state (where again $J=V_\mu=0$), we have 
$\hat h_{\fk{k}}(t)=\hat h_{\fk{k}}^{\rm out}\exp\{iUt/2\}$ and 
$\hat p_{\fk{k}}(t)=\hat p_{\fk{k}}^{\rm out}\exp\{-iUt/2\}$ once more. 
Consequently, initially and finally the positive or negative frequency 
components of the field 
$\hat\phi_{\fk{k}}^+(t)=\hat h_{\fk{k}}(t)+\hat p_{\fk{k}}(t)$ in 
Eq.~(\ref{scalar-field}) yield the initial and final particle or hole 
operators, respectively. 
As a result, a mixing of positive and negative frequencies of the 
$\hat\phi_+$-field in Eq.~(\ref{scalar-field}) as given by 
the Bogoliubov coefficients $\alpha_{\fk{k}}$ and $\beta_{\fk{k}}$
directly corresponds to a mixing of particle and hole operators 
\bea
\label{mixing}
\hat p_{\fk{k}}^{\rm out}
=
\alpha_{\fk{k}}\hat p_{\fk{k}}^{\rm in}+
\beta_{\fk{k}}\hat h_{\fk{k}}^{\rm in} 
\,.
\ea
In Fourier space, the initial conditions discussed after 
Eq.~(\ref{effective}) 
%
%(Mott state) 
%$\langle\hat h_\mu^\dagger\hat h_\nu\rangle_0=\delta_{\mu\nu}$,
%$\langle\hat h_\mu^\dagger\hat p_\nu\rangle_0=
%\langle\hat p_\mu^\dagger\hat h_\nu\rangle_0=
%\langle\hat p_\mu^\dagger\hat p_\nu\rangle_0=0$,
read 
$\langle\hat p_{\fk{k}}^\dagger\hat p_{\fk{k}}\rangle_{\rm in}=
\langle\hat p_{\fk{k}}^\dagger\hat h_{\fk{k}}\rangle_{\rm in}=
\langle\hat h_{\fk{k}}^\dagger\hat p_{\fk{k}}\rangle_{\rm in}=0$
and 
$\langle\hat h_{\fk{k}}^\dagger\hat h_{\fk{k}}\rangle_{\rm in}=1$.
Thus, inserting (\ref{mixing}), we get 
\bea
\langle\hat p_{\fk{k}}^\dagger\hat p_{\fk{k}}\rangle_{\rm out}
=
|\beta_{\fk{k}}|^2
\,.
\ea
This determines the on-site particle/hole probability via 
$\bra{2}\hat\rho_\mu\ket{2}_{\rm out}=
\langle\hat p_\mu^\dagger\hat p_\mu\rangle_{\rm out}
=\sum_{\fk{k}}|\beta_{\fk{k}}|^2/N
=\bra{0}\hat\rho_\mu\ket{0}_{\rm out}$
according to Eq.~(\ref{one-site}) and the
aforementioned particle-hole symmetry $f^{11}_{\mu\nu}=f^{22}_{\mu\nu}$. 
%
%\bea
%\bra{2}\hat\rho_\mu\ket{2}_{\rm out}
%=
%\langle\hat p_\mu^\dagger\hat p_\mu\rangle_{\rm out}
%=
%\frac{1}{N}\sum\limits_{\fk{k}}|\beta_{\fk{k}}|^2
%\,.
%\ea
%
%Note that the aforementioned particle hole symmetry implies  
%$\bra{0}\hat\rho_\mu\ket{0}_{\rm out}=\bra{2}\hat\rho_\mu\ket{2}_{\rm out}$.
%
As expected from the exact analogy between Eqs.~(\ref{Klein-Gordon}) and 
(\ref{scalar-field}), we infer the same exponential scaling as in the 
Sauter-Schwinger effect (\ref{probability}) 
%for small $\f{k}$
%
\bea
\label{beta}
|\beta_{\fk{k}}|^2
\propto 
%\approx  \frac{\pi^2}{36}
\exp\left\{-\pi\,
\frac{(\Delta{\cal E})^2/4+c^2_{\rm eff}\f{k}^2_\perp}{|\na V|c_{\rm eff}}
\right\}
\,,
\ea
for small $\f{k}$ and small and slowly varying lattice tilts. % $|\na V|$. 
Note that this expression is non-perturbative in $J$ and $U$, we only
exploited the $1/Z$-expansion.

\paragraph{Bloch oscillations.}

The exact analogy established above applies to small lattice tilts and 
large length scales. 
For large potential gradients, the long-wavelength expansion is not 
applicable anymore and the lattice structure becomes important. 
For simplicity, we assume a square lattice in the following, 
with the gradient pointing along a lattice axis, 
but the results can easily be generalized. 
For large tilts, one obtains resonance effects which can be understood 
by considering the Fourier transform of Eqs.~(\ref{f11}) with the 
time-dependent $T_{\fk{k}}$ from Eq.~(\ref{T_k}) 
\bea 
\label{f2}
(i\partial_t-U+3JT_{\fk{k}}) f_{\fk{k}}^{12}
&=&
-\sqrt{2}JT_{\fk{k}}(f_{\fk{k}}^{11}+f_{\fk{k}}^{22}+1)
\,,
\nn
i\partial_t f_{\fk{k}}^{11}=i\partial_t f_{\fk{k}}^{22}
&=&
\sqrt{2}JT_{\fk{k}}(f_{\fk{k}}^{12}-f_{\fk{k}}^{21}) 
\,,
\eea 
where $f_{\fk{k}}^{21}=(f_{\fk{k}}^{12})^*$.
For a constant gradient $\na V$, the $T_{\fk{k}}$ obey an oscillatory 
time-dependence due to the periodicity in $\f{k}\to\f{k}+t\na V$, 
which is the basis of the well-known Bloch oscillations \cite{Bloch,BlochTheo}. 
For simplicity, we first consider the limit of small $J$ which 
facilitates a perturbative solution of Eqs.~(\ref{f2}). 
To lowest oder in $J$, we obtain resonant growth if the potential
difference $\Delta V$ between two neighbouring lattice sites equals
the gap $\Delta{\cal E}=U+\ord(J)$
\bea
f_{\fk{k}}^{12}=\frac{i J t}{2\sqrt{2}}e^{-i U t}+\mathcal{O}(J^2)
\leadsto
f_{\fk{k}}^{11}=\frac{J^2 t^2}{8}+\mathcal{O}(J^3)
\,.
\ea
This resonance has a width of $\sqrt{2}J$ and corresponds to tunneling
to the nearest neighbouring site. 
In higher orders in $J$, we also obtain resonant tunneling to 
next-to-nearest neighbours for $2\Delta V=\Delta{\cal E}$ and so on.
In contrast to the non-perturbative result (\ref{beta}), this process
is more similar to electron-positron pair creation in the perturbative
multi-photon regime, see, e.g., \cite{Schwinger2}.

Beyond perturbation theory in $J$, we may employ Floquet analysis \cite{Floquet}
to find the various resonance bands which yield an exponential growth 
of the solutions to Eqs.~(\ref{f2}). 
The associated Floquet exponent for the first resonance at 
$\Delta V=\Delta{\cal E}$ reads $J/(\sqrt{2}\Delta V)$ and for 
the second resonance $2\Delta V=\Delta{\cal E}$, 
it is $3J^2/(4\sqrt{2}[\Delta V]^2)$ etc.

\paragraph{Conclusions.}

In summary, we demonstrated that electron-positron pair production by a 
strong (and slowly varying) electric field and particle-hole pair creation 
in a slightly tilted Mott insulating Bose-Hubbard lattice are governed 
by the {\em same equations.} 
This quantitative analogy is sketched in the following table: 

\bigskip

\begin{tabular}{|c|c|}
\hline
Sauter-Schwinger effect & \;Bose-Hubbard model\; \\ 
\hline
electrons \& positrons & particles \& holes \\ 
Dirac sea & Mott state \\ 
\;mass of electron/positron\; & energy gap $\Delta{\cal E}$ \\
electric field $\f{E}$ & lattice tilt $V_\mu$ \\ 
speed of light $c$ & velocity $c_{\rm eff}$ \\ 
\hline
\end{tabular}

\bigskip

This analogy allows us to apply the vast machinery developed for the 
Sauter-Schwinger effect \cite{Schwinger1,Schwinger2,Schwinger3} to condensed matter theory. 
For example, small time-dependent variations of the potential gradient 
may significantly enhance the tunneling probability \cite{Schwinger3} 
even if the rate of change is much smaller than the energy gap 
$\Delta{\cal E}$.
These effects should be observable with atoms in optical lattices \cite{BoseHubbard}
where the experimental technique now even allows the in situ detection 
of single atoms \cite{InSitu}.
%
%Our results should also be related to recent investigations of the 
%onset of dielectric breakdown [...] studied in fermionic Hubbard model.  
%
Finally, it would be interesting to study whether this quantitative analogy 
can be extented to the onset of dielectric breakdown in the fermionic 
Hubbard model, see, e.g., \cite{DieBreak}. 

\begin{acknowledgments}
This work was supported by the SFB/TR 12 of the 
German Research Foundation (DFG).
Helpful discussions with K.~Krutitsky are gratefully acknowledged.
\end{acknowledgments}%

$^*$ {\tt ralf.schuetzhold@uni-due.de}

\bibliographystyle{apsrmp4}

\begin{thebibliography}{99}
\bibitem{Schwinger1}
F.~Sauter, Zeits f.~Physik {\bf 69}, 742 (1931);
%Die Reflexion von Elektronen an einem Potentialsprung nach der relativistischen Dynamik von Dirac
J.~Schwinger, Phys.~Rev.~{\bf 82}, 664 (1951);
%On Gauge Invariance and Vacuum Polarization
L.~V.~Keldysh, Sov.~Phys.~JETP {\bf 02}, 1307 (1965).

\bibitem{Schwinger2}
E.~Brezin and C. Itzykson, Phys.~Rev.~D {\bf 2}, 1191 (1970);
%Pair Production in Vacuum by an Alternating Field
G.~V.~Dunne, Eur.~Phys.~J.~D, {\bf 55}, 327 (2009).
%New strong-field QED effects at extreme light infrastructure

\bibitem{Schwinger3}
R.~Sch\"utzhold, H.~Gies, and G.~Dunne, Phys. Rev. Lett. {\bf 101}, 130404 (2008).
%Dynamically Assisted Schwinger Mechanism

\bibitem{BoseHubbard}
M.~Greiner, O.~Mandel, T.~Esslinger, T.~W.~H\"{a}nsch et al., Nature {\bf 415}, 39 (2002);
%Quantum phase transition from a superfluid to a Mott insulator in a gas of ultracold atoms
J.~Zakrzewski, Phys.~Rev.~A {\bf 71}, 043601 (2005);
%Mean-field dynamics of the superfluid-insulator phase transition in a gas of ultracold atoms
I.~Bloch, Nature Physics, {\bf 1}, 23 (2005);
%Ultracold quantum gases in optical lattices
C.~Sias, A.~Zenesini, H.~Lignier, S.~Wimberger et al., Phys.~Rev.~Lett. {\bf 98}, 120403 (2007);
%Resonantly Enhanced Tunneling of Bose-Einstein Condensates in Periodic Potentials
M.~Raizen, C.~Salomon, and Q.~Niu, Physics Today, {\bf 50}, 30 (1997).
%New Light on Quantum Transport

\bibitem{Bloch}
O.~Morsch, J.~H.~M\"uller, M.~Cristiani, D.~Ciampini et al., Phys.~Rev.~Lett. {\bf 87}, 140402 (2001);
%Bloch Oscillations and Mean-Field Effects of Bose-Einstein Condensates in 1D Optical Lattices
Phys.~Rev.~A {\bf 65}, 063612 (2002);
%Experimental properties of Bose-Einstein condensates in one-dimensional optical lattices: Bloch oscillations, Landau-Zener tunneling, and mean-field effects
M.~B.~Dahan, E.~Peik, J.~Reichel, Y.~Castin et al., Phys.~Rev.~Lett. {\bf 76}, 4508 (1996).
%Bloch Oscillations of Atoms in an Optical Potential



\bibitem{NavSch}
P.~Navez and R.~Sch\"utzhold, Phys.~Rev.~A {\bf 82}, 063603 (2010).
%Emergence of coherence in the Mott-insulator--superfluid quench of the Bose-Hubbard model

\bibitem{hierarchy} 
R.~Kubo, J.~Phys.~Soc.~Japan {\bf 17}, 1100 (1962);
R. Balescu, {\it Equilibrium and Nonequilibrium
Statistical Mechanics} (Wiley, New York, 1975).


\bibitem{BlochTheo}
D.~Witthaut, M.~Werder, S.~Mossmann, and H.~J.~Korsch, Phys.~Rev.~E {\bf 71}, 036625 (2005);
%Bloch oscillations of Bose-Einstein condensates: Breakdown and revival
S.~Sachdev, K.~Sengupta, and S.~M.~Girvin, Phys.~Rev.~B {\bf 66}, 075128 (2002);
%Mott insulators in strong electric fields
S.~Wimberger, R.~Mannella, O.~Morsch, E.~Arimondo et al., Phys.~Rev.~A {\bf 72}, 063610 (2005);
%Nonlinearity-induced destruction of resonant tunneling in the Wannier-Stark problem
A.~R.~Kolovsky and H.~J.~Korsch, Phys.~Rev.~A {\bf 67}, 063601 (2003);
%Bloch oscillations of cold atoms in two-dimensional optical lattices
Int.~J.~Mod.~Phys.~B {\bf 18}, 1235 (2004);
%Bloch oscillations of cold atoms in optical lattices
A.~R.~Kolovsky, Phys.~Rev.~A {\bf 70}, 015604 (2004);
%Bloch oscillations in the Mott-insulator regime
Phys.~Rev.~Lett. {\bf 90}, 213002 (2003);
%New Bloch Period for Interacting Cold Atoms in 1D Optical Lattices
A.~R.~Kolovsky and A.~Buchleitner, Phys.~Rev.~E {\bf 68}, 056213 (2003);
%Floquet-Bloch operator for the Bose-Hubbard model with static field
A.~R.~Kolovsky, H.~J.~Korsch and E.-M.~Graefe, Phys.~Rev.~A {\bf 80} 023617 (2009);
%Bloch oscillations of Bose-Einstein condensates: Quantum counterpart of dynamical instability
D.-I.~Choi and Q.~Niu, Phys.~Rev.~Lett. {\bf 82}, 2022 (1999).
%Bose-Einstein Condensates in an Optical Lattice


\bibitem{Floquet}
Z.~X.~Wang and D.~R.~Guo, {\it Special Functions} (World Scientific, 1989).

\bibitem{InSitu}
P.~W\"urtz, T.~Langen, T.~Gericke, A.~Koglbauer et al., Phys.~Rev.~Lett. {\bf 103}, 080404 (2009);
%Experimental Demonstration of Single-Site Addressability in a Two-Dimensional Optical Lattice
W.~S.~Bakr, A.~Peng, M.~E.~Tai, J.~Simon et al., Science {\bf 329}, 547 (2010);
%Probing the superfluid-to-Mott insulator transition at the single-atom level
N.~Gemelke, X.~Zhang, C.-L.~Hung, and C.~Chin, Nature {\bf 460}, 995 (2009);
%In situ observation of incompressible Mott-insulating domains in ultracold atomic gases
J.~F.~Sherson, C.~Weitenberg, M.~Endres, M.~Cheneau et al., Nature {\bf 467}, 68 (2010).
%Single-atom-resolved fluorescence imaging of an atomic Mott insulator

\bibitem{DieBreak}
M.~Eckstein, T.~Oka, and P.~Werner, Phys.~Rev.~Lett. {\bf 105}, 146404 (2010);
%Dielectric Breakdown of Mott Insulators in Dynamical Mean-Field Theory
T.~Oka, R.~Arita, and H.~Aoki, Phys.~Rev.~Lett. {\bf 91}, 066406 (2003);
%Breakdown of a Mott Insulator: A Nonadiabatic Tunneling Mechanism
T.~Oka and H.~Aoki, Phys.~Rev.~B {\bf 81}, 033103 (2010).
%Dielectric breakdown in a Mott insulator: Many-body Schwinger-Landau-Zener mechanism studied with a generalized Bethe ansatz

\end{thebibliography}

\end{document}